\documentclass[12pt,epsf]{article}
\usepackage{graphicx, amsmath}

\begin{document}
\sloppy
\begin{flushright}{SIT-HEP/TM-59}
\end{flushright}
\vskip 1.5 truecm
\centerline{\large{\bf Spotted inflation}}
\vskip .75 truecm
\centerline{\bf Tomohiro Matsuda\footnote{matsuda@sit.ac.jp}}
\vskip .4 truecm
\centerline {\it Laboratory of Physics, Saitama Institute of Technology,}
\centerline {\it Fusaiji, Okabe-machi, Saitama 369-0293, 
Japan}
\vskip 1. truecm
\makeatletter
\@addtoreset{equation}{section}
\def\theequation{\thesection.\arabic{equation}}
\makeatother
\vskip 1. truecm
\begin{abstract}
\hspace*{\parindent}
We describe new scenarios for generating curvature
 perturbations when inflaton (curvaton) has significant
 interactions. 
We consider a ``spot'', which arises from interactions associated
 with an enhanced symmetric point (ESP) on the trajectory.
Our first example uses the spot to induce a gap in the field equation.
We observe that the gap in the field equation may cause generation
 of curvature perturbation if it does not appear simultaneous in 
 space.
The mechanism is similar to the scenario of inhomogeneous phase
transition.
Then we observe that the spot interactions may initiate warm inflation
in the cold Universe.
Creation of cosmological perturbation is discussed in
relation to the inflaton dynamics and the modulation associated with the spot
interactions.
\end{abstract}

\newpage
\section{Introduction}
Inflation is an important cosmological event in the early Universe,
which solves many cosmological problems and leads to generation of
cosmological perturbations.  
Besides the standard scenario of single-field inflation, which
(basically) generates Gaussian perturbation, scenarios using multiple
fields for creating cosmological perturbations is very interesting
because of their rich variety and the possibility of generating non-standard
spectrum.
For instance, long-wavelength inhomogeneity of a light scalar field
 in a decoupled sector may give rise to superhorizon fluctuation of
 couplings or masses in the low-energy effective
 action\cite{modulation-original}, which may lead to another source
 of cosmological perturbation.
Models with light-field fluctuations are called
``modulated''\cite{modulation-original}  or
``inhomogeneous''\cite{IH-R, IH-pt} scenarios. 
Cosmological phase transitions, which may include conventional
reheating after inflation (e.g., ``inhomogeneous reheating'' 
considered in Ref.\cite{IH-R}), may 
then not occur simultaneous in space , but may occur with time 
lags in different Hubble patches due to the long-wavelength
inhomogeneity of the model parameters.\footnote{An interesting model
of inhomogeneous phase transition has been considered in
Ref.\cite{IH-pt}, which shows that cosmological perturbations may be
created at the electroweak phase transition. 
The result also shows that phase transitions may be a generic source of
non-Gaussianity.} 
In Section \ref{gapin} of this paper we first apply the idea of inhomogeneous
transition to the gapped equation of the inflaton field.
In this scenario, the gap in the field equation arises from
the interactions at an enhanced symmetric point(ESP).
Since the interactions at the ESP may cause instant preheating,
the cold Universe may turn into a warm one after the inflaton
passes through the ESP.
Except for warm inflation and thermal inflation, the Universe during
inflation is usually presumed
to be cold because of the rapid red-shifting of the primeval 
radiation.\footnote{Note however the Hawking temperature is intrinsic to
de Sitter space. In this sense, the Universe during inflation is
cold but the temperature is not supposed to be zero.}
Reheating after inflation is thus essential for the cold inflationary
scenario in order to recover the standard hot Universe.
However, since the high-temperature reheating may cause thermal production of
unwanted relics, entropy production after reheating (e.g, secondary inflation
that leads to mild reheating) is in some cases useful for
ensuring successful nucleosynthesis.  
A significant example of this kind is thermal
inflation\cite{thermal-original}, in which the radiation energy
decreases rapidly during thermal inflation and the trapping of the
inflaton field is caused by the symmetry restoration.
Besides thermal inflation, which is not suitable for primary inflation,
the warm inflation scenario\cite{warm-inflation-original} also uses 
radiation during inflation.
What most notably differentiates warm inflation from thermal inflation
is the evolution of the radiation during inflation.
In warm inflation, the temperature does not decrease rapidly
 but is sustained by the dissipation from the inflaton motion.
Moreover, the inflaton field in the warm inflation scenario is not
trapped in the false vacuum but rolls slowly during inflation, where the
frictional force is enhanced by the dissipation of the kinetic energy.
In Section \ref{ex2}, using the interactions of the ESP, we consider a new 
hybrid of thermal and warm inflation in which slow-roll is due to the
dissipation of the field.
We will show that the scenario causes generation of curvature
perturbation if it is combined with inhomogeneous transition\cite{IH-pt}.
This point is in contrast to the conventional thermal inflation model,
in which generation of significant curvature perturbation is 
impossible.\footnote{However, utilizing a fluctuating
coupling of a light field, it has been shown in
Ref.\cite{thermal-curvgen} that the mechanism of inhomogeneous phase 
transition\cite{IH-pt} may cause generation of curvature perturbation
at the end of thermal inflation, which may be used to liberate some
inflation models. Namely, if the curvature perturbation is generated at  
secondary inflation, primary inflation is free from constraints
related to the curvature perturbation.}  
The spotted inflation model considered in Section \ref{ex2} uses
dissipation for the slow-roll. 
The idea is basically the same as the dissipative curvaton model in 
Ref.\cite{Dissipative-curvatons}, but in the present paper 
it will be shown that the physics related to the shifted ESP interactions 
leads to significant differences that differentiate the spotted model
from the original model of dissipative curvaton.
In this paper spotted inflation is basically considered for the scenario of
primary inflation.
However, it can be used to realize a curvaton-like scenario, which appears
as secondary inflation.
The curvaton-like scenario of spotted inflation (spotted curvatons) 
will be mentioned briefly in Section \ref{curv}. 

This paper considers ``spot'' on the field trajectory, which
raises significant interactions that affect the field motion.
The ``spot'' is defined in Section \ref{define-spot} 
and in Fig. \ref{fig:spot}.  
Consequences of the spot interactions will be discussed in relation to
the possible creation of cosmological perturbations.
The purpose of this paper is to show some useful examples, which
give intuitive explanations for why interactions of the fields
are potentially important for discussing cosmological events.

\section{Inhomogeneous transition with velocity gap}
\label{gapin}
The scenario of inhomogeneous phase transition \cite{IH-pt} uses a
 gap $\Delta  n$ that appears in the scaling relation $\rho\propto
 a^{-n}$, where  $\rho$ and $a$ are the energy 
 density and the scale of the Universe, respectively.
Then the gap in the scaling relation causes generation of the curvature
 perturbation when it does not appear simultaneous in space.
In contrast to the original scenario of inhomogeneous reheating, the
present scenario uses gap in the inflaton velocity; $\Delta \dot{\phi}\ne 0$.
We describe creation of curvature perturbation 
at the modulated (inhomogeneous) boundary of the transition,
first using a toy model and then using spot interactions.

\subsection{Toy model}
\begin{figure}[h]
 \begin{center}
\begin{picture}(200,220)(100,360)
\resizebox{15cm}{!}{\includegraphics{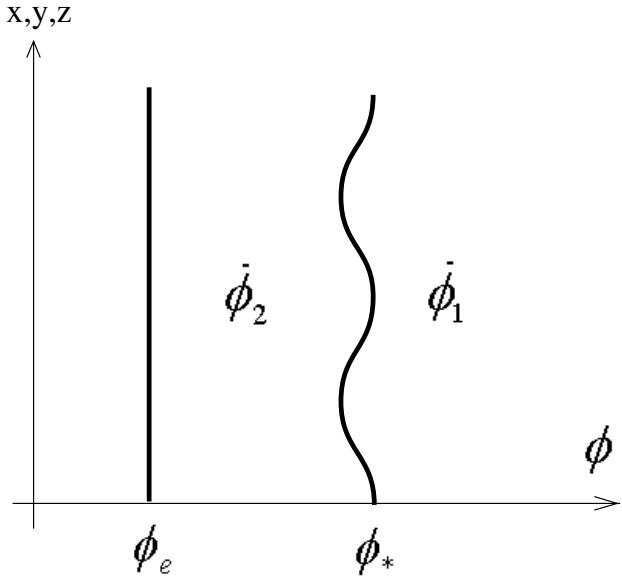}} 
\end{picture}
 \caption{There are two distinct regions of the inflaton trajectory, 
in which inflaton slow-roll velocities are given by $\dot{\phi}_1$ and 
$\dot{\phi}_1$.
 The boundary that separates the two regions is $x,y,z$-dependent and 
appears inhomogeneous in space.
In the above picture slow-roll inflation ends at $\phi_e$.} 
\label{fig:basic}
 \end{center}
\end{figure}

Our first model is very simple.
It contains the minimum ingredients for describing the
 mechanism of generating metric perturbations using the inhomogeneous
 boundary.
The boundary is associated with the gap in the inflaton velocity.
We assume two distinct regions of the inflaton trajectory, where the
 inflaton velocities are unique.
These velocities are denoted by $\dot{\phi}_1$ and
 $\dot{\phi}_2$. 
At the boundary ($\phi=\phi_*$) we consider a gap 
$\Delta \dot{\phi}\equiv \dot{\phi}_2-\dot{\phi}_1> 0$ for the inflaton
velocity.\footnote{We implicitly assume that there is a gap in the
inflaton potential at $\phi=\phi_*$.}
The situation is shown in Fig.\ref{fig:basic}, where the inflaton moves
from right to left.
Although the gap itself cannot be the seed of the cosmological
 perturbations of the superhorizon scale, inhomogeneity of the
 boundary ($\delta \phi_*\ne 0$) may cause
 curvature perturbations. 
In fact, long-wavelength inhomogeneity of light scalar fields
 in a decoupled sector may give rise to superhorizon fluctuations of
 couplings or masses in the low-energy effective action (this is 
the scenario called  ``modulation''\cite{modulation-original}),
 which may cause inhomogeneity  ($\delta \phi_*\ne 0$) at the boundary.
Cosmological transition over the boundary, which accompanies $\Delta
 \dot{\phi}\ne 0$ and $\delta \phi_* \ne 0$ at the same time, 
then may not occur simultaneous in space but with time lags in
 different Hubble patches.
As a result, the inhomogeneity causes $\delta t$ at the transition,
which is estimated as
\begin{equation}
\delta t \simeq \delta \phi_* \left(\frac{1}{\dot{\phi}_1}
-\frac{1}{\dot{\phi}_2}\right)=
\frac{ \delta \phi_*}{\dot{\phi}_1}
\left(1-\frac{\dot{\phi}_1}{\dot{\phi}_2}\right).
\end{equation}
Using the $\delta N$ formalism, this result leads to
\begin{equation}
\label{basic-deltaN}
\delta N \simeq 
H \frac{\delta \phi_*}{\dot{\phi}_1}
\left(1-\frac{\dot{\phi}_1}{\dot{\phi}_2}\right),
\end{equation}
which shows that the gap appearing in the inflaton velocity
may cause $\delta N\ne 0$ during inflation.
In this model $\delta N \ne 0$ is possible even if
the initial amplitude of the inflaton perturbation at the horizon exit
(in this paper the amplitude at the horizon exit is denoted 
by the subscript ``k'', such as $\delta \phi_k$ for the field $\phi$) is
vanishing.
The generation of $\delta N$ is possible if the boundary of the
transition $\dot{\phi}_2\rightarrow \dot{\phi}_1$ is inhomogeneous
 due to modulation.\footnote{The
 most significant model of this kind is found in
 Ref.\cite{At-the-end-lyth,At-the-end-kov}, where the scenario for generating 
cosmological perturbations at the end of hybrid
inflation is discussed.
In this case, the inflaton velocity changes suddenly at $\phi=\phi_e$
 due to the instability of the waterfall field.
The scenario uses $\delta \phi_e \ne 0$, which is caused by the additional
inflaton field\cite{At-the-end-lyth} or by modulation\cite{At-the-end-kov}.}  
The typical length scale of the cosmological perturbation is determined
by the time for horizon exit, when the related perturbation (perturbation of
the massless degree of freedom) is frozen. 
To show a more concrete example, let us consider the specific case in which 
$\phi_*$ is given by the function of light fields $\sigma^{(i)}$,
(i=1,2,...n).  
Then considering amplitudes $\delta \sigma^{(i)}_k$, which exit
the horizon at the beginning of inflation,
the inhomogeneity of the boundary is expressed as
\begin{equation}
\delta \phi_* \simeq \sum_i
\frac{\partial \phi_*}{\partial \sigma^{(i)}}
\delta \sigma_k^{(i)}
+...,
\end{equation}
where higher perturbations are disregarded for simplicity.
The amplitude of $\delta \phi_*$ can be much larger than $\delta \phi_k$.
Namely, $\delta \phi_*\gg \delta \phi_k \simeq \delta \sigma_k^{(i)}$ is
acceptable in the modulation scenario.

\subsection{Spotted inflation (non-dissipative)}
\label{define-spot}
\begin{figure}[h]
 \begin{center}
\begin{picture}(200,180)(120,40)
\resizebox{15cm}{!}{\includegraphics{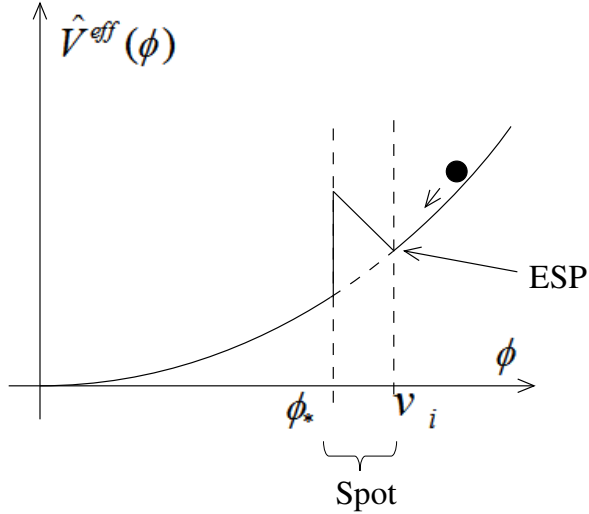}} 
\end{picture}
 \caption{Creation of $\chi_i$ particles causes a significant gap
  in the effective potential until the particles decay into light
  fermions.
Note that the ``spot'' is not defined by the point.
The definition of the spot is different from the ESP.} 
\label{fig:spot}
 \end{center}
\end{figure}

In this section we consider a simple scenario in which the
inflaton field $\phi$ rolls slowly during cold inflation ($\epsilon <1$).
At the enhanced symmetry point, although the inflaton is still slow-rolling,
the inflaton may dump its kinetic energy into the production of other
particles $\chi_i$.
This leads to the event called preheating\cite{PR-original}, although
this event does not lead to reheating in the conventional
sense.
The couplings are described by the interactions of the form
\begin{equation}
\sum_i \left[\frac{1}{2}g_i^2(\phi-v_i)^2 \chi^2_i + h_i \chi_i 
\overline{\psi}\psi\right].
\end{equation}
We assume for simplicity that $\chi_i$ particles (preheat fields) 
do not have bare
masses other than the interactions in the above formula.
The slow-roll equation suggests that during inflation the inflaton
moves with velocity $\dot{\phi}_0 \simeq \sqrt{\epsilon}
HM_p$, where
$\epsilon\equiv \frac{M_p^2}{2} \left(\frac{V'}{V}\right)^2$ is the
conventional slow-roll parameter. 
The adiabatic condition is violated and particle production occurs when
$\dot{m}_{\chi_i}(\sim g_i|\dot{\phi}_0|)$ 
becomes greater than $m_{\chi_i}^2(\sim g_i^2 |\phi-v_i|^2)$.
This condition defines the region for the efficient particle production given by
\begin{equation}
 |\phi-v_i|^2<\frac{\dot{\phi}_0}{g_i}\sim 
\frac{\sqrt{\epsilon}HM_p}{g_i}.
\end{equation}
When the inflaton field $\phi$ passes through the ESP at $v_i$, the
corresponding particle $\chi_i$ becomes light and is produced with 
number density $n_i$, which is maximally\cite{PR-original}
\begin{equation}
n_i^{Max} \sim \frac{g_i^{3/2}}{(2\pi)^3}
\left(\dot{\phi}_0\right)^{3/2}.
\end{equation}
$n_i$ may redshift with the Universe expansion, however
the process of particle production occurs within the time
\begin{equation}
\Delta t \sim \frac{\sqrt{\dot{\phi}_0/g_i}}{\dot{\phi}_0}\sim 
\frac{1}{\sqrt{g_i\dot{\phi}_0}}\sim \frac{1}{g^{1/2}_i(2\epsilon)^{1/4}
\sqrt{HM_p}},
\end{equation}
which is much smaller than the age of the Universe if 
$H\ll g_i\sqrt{2\epsilon}M_p$.
We consider the case with $\Delta t << H^{-1}$.
The energy density of $\chi_i$ after preheating is given by
$\rho_i \simeq g_i|\phi-v_i|n_{\chi_i}$.
Considering $\rho_i$, the effective
$\phi$ equation of motion after the particle production  is  given by
\begin{equation}
\ddot{\phi}+3H\dot{\phi}+V' - g_i n_i=0.
\end{equation}
The last term determines the inflaton motion when $|V'| < |g_i n_i|$, which
occurs if
\begin{equation}
\label{gap-eq1}
V^{1/4}<\frac{g_i^{5/2}}{(2\pi)^3 3^{3/4}}(2\epsilon)^{1/4} M_p
\sim 10^{-3}g_i^{5/2}\epsilon^{1/4} M_p,
\end{equation}
where $V'$ and $H$ are replaced using $V'= \sqrt{2\epsilon}V/M_p$
and $H^2 = V/3M_p^2$.
In the above equation we considered $n_i\simeq n^{max}_i$, but the
assumption is not essential for the analysis.
Our scenario works when the condition (\ref{gap-eq1}) is satisfied by the
inflation model.
The slow-roll parameter $0<\epsilon<1$ in the above equation is defined
at the ESP, which may be different from the one at the horizon.
The scenario leads to a simple model in which there is a gap in
the inflaton equation in the midst of the inflationary epoch.
A schematic picture for the situation is shown in
Fig.\ref{fig:spot}. 
We consider the approximation that the gap in the field equation
($\sim - g_i n_i$) 
appears suddenly at $\phi\simeq v_i$, 
where $\chi_i$ particles are created, and then disappears at $\phi_*$,
where $\chi_i$ with the mass $m_{\chi_i}\sim g_i |\phi-v_i|$ decays with
the decay rate 
\begin{equation}
\Gamma_i(\chi_i\rightarrow 2\psi)\simeq 
\frac{h_i^2}{8\pi} m_{\chi_i}\simeq 
\frac{h_i^2}{8\pi} g_i |\phi-v_i|.  
\end{equation}
We thus define the ``spot'', which starts at $\phi=v_i$ and ends at
$\phi=\phi_*$.
The spot has the required gap $\Delta \dot{\phi}\ne 0$
at $\phi=\phi_*$. 
The end boundary of the spot is defined using $\Gamma_i \sim H$,
which leads to
\begin{equation}
\phi_*\simeq v_i-\frac{8\pi}{h_i^2 g_i}  H.
\end{equation}
Defining the effective potential $\hat{V}(\phi)^{eff}$ during
$\phi_*<\phi<v_i$, which is given by
\begin{equation}
\hat{V}(\phi)^{eff}\equiv V(\phi)+g_i|\phi-v_i|n_{\chi_i},
\end{equation}
we consider the inflaton field equation that has an additional term in the
effective potential.
Considering the modest approximation that $n_i$ is nearly constant during
$\phi_*<\phi< v_i$,
which is conceivable because the process occurs in a short time period,
the inflaton velocity decreases with constant acceleration(deceleration).
Then the inflaton velocity can be expressed as 
\begin{equation}
\dot{\phi}(t)\simeq \dot{\phi}_0+g_i n_i t,
\end{equation}
where $t=0$ is defined at $\phi=v_i$ and the initial velocity is
 $\dot{\phi}(0)\simeq \dot{\phi}_0$.
The important assumption in our scenario is that $\chi_i$
particles decay after preheating, which  
occurs before the inflaton makes a turn.\footnote{The
scenario describing the inflaton that makes a turn {\bf before} 
the decay is called trapped inflation\cite{Trapped-Inf}.}
Since the deceleration is almost constant after the ESP, 
the condition for the inflaton not to make a turn before the
decay of $\chi_i$ particles is given by
\begin{equation}
\frac{1}{2}\frac{\dot{\phi}_0^2}{g_i n_i}>
 |\phi_*-v_i|,
\end{equation}
which leads to
\begin{equation}
h_i^2 > (h_i^c)^2\simeq 0.1 \frac{\sqrt{g_i}}{\epsilon^{1/4}}
\sqrt{\frac{H}{M_p}}.
\end{equation}
The most significant effect appears when 
$h_i^2 \simeq (h_i^c)^2$,
which leads to $\dot{\phi}_1/\dot{\phi}_2\ll 1$ in
Eq.(\ref{basic-deltaN}). 

In this work we consider the perturbation $\delta \phi_*\ne 0$ of the
boundary at $\phi=\phi_*$, which can be caused by the modulation of the 
couplings $h_i$.
Here we consider the function $h_i(\sigma^{(j)})$ for $h_i$, where
the origin of the primordial perturbation is $\delta \sigma_k^{(j)} \ne
0$ of the light field $\sigma^{(j)}$.
We have in mind the situation that the inflaton velocity decreases
after preheating and finally reaches 
$\dot{\phi_e}\ll\dot{\phi}_0$ at $\phi=\phi_*$.
The inflaton recovers $\dot{\phi}\simeq \dot{\phi}_0$ as soon as it exits
the spot.\footnote{The decay products of the $\chi_i$ particles may
thermalize to cause significant dissipation to the inflaton motion.
However, the thermalization process depends crucially on the
interactions of the radiation sector.
If the decay products thermalize and cause significant dissipation of
the inflaton motion, the inflaton slow-roll velocity is given by
$\dot{\phi}_0=V'/3H(1+r)$. 
This idea is examined in the next section
for dissipative inflation, where the ratio $r$ is defined.
In this section we consider the cold inflation scenario, in which thermalization
and dissipation are not significant for the inflaton motion.}
Since the whole process occurs within a short-time period compared with the age
of the Universe, expansion of the Universe and the dilution of
$n_{\chi_i}$ are disregarded.

Assuming modulation given by
\begin{equation}
\delta \phi_*\simeq \frac{16\pi}{h_i^2g_i}\frac{\delta h_i}{h_i}H,
\end{equation}
the inhomogeneous transition at $\phi=\phi_*$ causes cosmological
perturbation.
Considering the approximation that the inflaton velocity changes
suddenly at $\phi=\phi_*$, the perturbation of
Eq.(\ref{basic-deltaN}) is expressed as
\begin{equation}
\delta N \simeq 
\frac{16\pi^2}{h_i^2g_i}\frac{\delta h_i}{h_i}
\left(1-\frac{\dot{\phi_e}}{\dot{\phi}_0}\right)\times
\left[\frac{H^2}{\dot{\phi}_0}\right].
\end{equation}

\section{Spotted inflation (dissipative)}
\label{ex2}
The role played by the spot may be more significant if it appears in 
warm inflation scenario.
In warm inflation the temperature during inflation is
$T>H$, which characterizes the  scenario.
During warm inflation the radiation is sustained by the
dissipation, which may lead to strong dissipation (SD) scenario
in which slow-roll inflation is realized with steep
potential.

In this section we consider a new scenario of inflation, in which
dissipative motion starts when the radiation is created by the decay of
the particle $\chi_i$.
The radiation sector is supposed to have strong interactions that are
needed for the rapid thermalization. 
This scenario of spotted inflation describes new aspects of warm
inflation associated with the spot interactions.

Significant evolution of the temperature is unavoidable for
the scenario, because the temperature at the beginning of dissipative
inflation is not the same as the ``equilibrium'' temperature.
Since the evolution is similar to thermal inflation,
we call this phase ``thermal phase'' in order to
distinguish this peculiar phase from the conventional
(``equilibrium'') phase of warm inflation.\footnote{Here we use the word
``equilibrium'' for the state that satisfies the specific condition
$\dot{\rho}_R\ \ll 4H\rho_R$.
 For the calculation of the dissipation coefficient we need to consider
 a small (or large) leave from the thermal equilibrium. 
The word ``equilibrium'' in this paper is used for the specific
condition, which should be distinguished from the ``thermal equilibrium''.} 
In this paper inflation with significant dissipation is called
``dissipative inflation'', and the ``equilibrium'' phase of dissipative
inflation is called ``warm phase''.\footnote{See
ref.\cite{Dissipative-curvatons} for another arugument for the
dissipative model, in which ESP is placed at the origin.}
As the result, in this paper the standard warm inflation phase is called
``dissipative inflation in the warm phase''.\footnote{Warm inflation is
defined to be any model where entropy is significant during inflation.
Since warm inflation models usually satisfies the ``equilibrium''
condition, we have to introduce the word ``dissipative inflation'' for
clarification.
This does not mean that warm inflation only applies when close to
``equilibrium''. } 

\begin{figure}[h]
 \begin{center}
\begin{picture}(200,220)(100,20)
\resizebox{15cm}{!}{\includegraphics{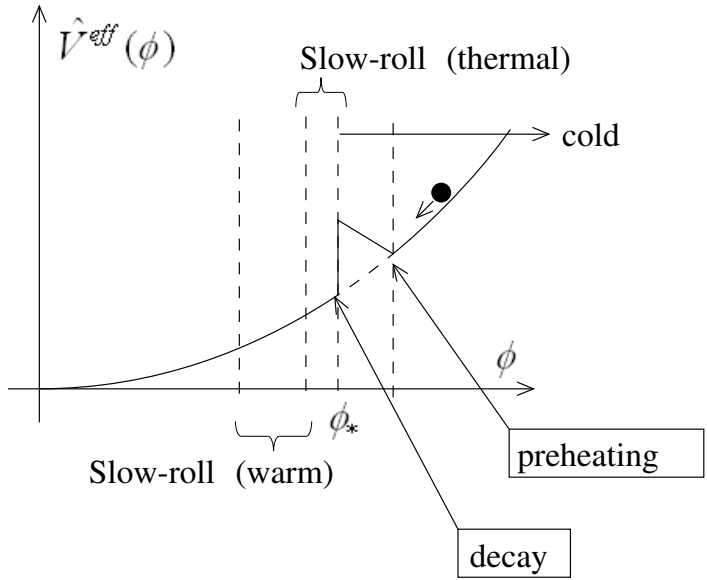}} 
\end{picture}
 \caption{There are two distinctive regions for the dissipative
  slow-roll.
In the first region, which is denoted by ``thermal'', radiation energy
  scales as $\rho_R\propto a^{-4}$ because the temperature is still 
larger than the ``equilibrium''.
In the second region, which is denoted by ``warm'', radiation is sourced
  by the dissipation and is sustained at $T_{eq}$.
What happens in the warm phase is the same as usual warm inflation.}
\label{fig:spot-warm}
 \end{center}
\end{figure}

The situation near the spot is explained in Fig.\ref{fig:spot-warm}.
Dissipative inflation caused by the spot may be short (i.e., the number of
e-foldings associated with the spotted inflation can be $N_e
\ll 60$).
Therefore, spotted inflation can be considered as a model of
secondary inflation that leads to mild reheating. 
The scenario of spotted inflation as secondary inflation is
mentioned separately in section \ref{curv}, where dissipative inflation
is considered for curvaton-like scenario.  

Before discussing the details, it is useful to
show brief outlook of the scenario.
At the ESP, which initiates the spot on the inflaton trajectory, 
the inflaton field $\phi$ causes preheating in the cold Universe.
The field $\phi$ may or may not be slow-rolling before
preheating.\footnote{We have in mind the situation that has been
discussed for trapped inflation scenarios\cite{Trapped-Inf}. 
The field $\phi$ starts to (slow/fast) roll in the cold Universe and passes
through the ESPs.
What discriminates our scenario from trapped inflation is that after
preheating at $\phi=v_i$ the same field $\phi$ successively 
produces dissipative inflation in the thermal background.}
Dissipative inflation starts with the thermal phase, as is shown in
Fig. \ref{fig:spot-warm}.
Then the warm phase may appear if the dissipation and thermalization 
after preheating is strong enough to sustain the radiation.
The mechanism of warm inflation and the conditions for the dissipative
slow-roll is discussed in Section \ref{diss-slo} .

For the dissipative inflation scenario we
consider dissipation during $\dot{m}_{\chi_i}<m_{\chi_i}^2$.
The condition is in contrast to the preheating condition at the ESP,
and it leads to   
\begin{equation}
 |\phi-v_i| > \sqrt{\frac{\dot{\phi}}{g}}.
\end{equation}
Dissipation starts after $\chi_i$ particles decay into radiation.
Let us remember that the origin of radiation at the beginning of warm
inflation is not clear in usual warm inflation scenario.
Of course, one may assume that warm inflation begins in the very hot
Universe, or that the radiation is created by the quasiparticle
dissipation that may be significant at zero-temperature.
However, another interesting story is that the cold Universe 
is turned into warm inflationary phase when significant radiation is
generated at some cosmological event.
In this work we introduce ``spot'', which realizes the scenario in which
radiation is created by the event (preheating) during inflation.

In this work we consider the scenario in which the beginning of
warm inflation is explained by the instant preheating at the spot.
We have in mind the situation that the radiation created at the spot is
large enough to initiate warm inflation, and the warm background 
($T\gg H$) is sustained by the subsequent dissipation of the field $\phi$. 
The inflaton potential may be steep, as far as it allows slow-roll in the
context of dissipative inflation.
The interactions of the spot are used to explain the subsequent dissipation.
One thing that must be considered with care is that 
the temperature just at the beginning of dissipative inflation is not
constant but scales as $\rho_R\propto a^{-4}$ until it reaches the
equilibrium value of the warm phase.
In the thermal phase, physics related to the dissipative slow-roll is
much different from what happens in the usual scenario of warm
inflation. 
The number of e-foldings elapsed in the thermal phase is given by
\begin{equation}
\label{n-th}
N_{th}\simeq \ln\frac{T_{*}}{T_{eq}},
\end{equation}
where $T_{*}$ and $T_{eq}$ denote the temperature at $\phi=\phi_*$
(i.e., the temperature of the radiation created by the spot)
and the equilibrium temperature of the warm inflationary phase
(i.e., the temperature when the thermal phase ends).
The process toward the equilibrium is fast, but it can cause inhomogeneity
$\delta N_{th}\sim (\delta T_*/T_*)-(\delta T_{eq}/T_{eq})$.
A significant sources of $\delta T_*$ is the inhomogeneous
preheating\cite{IH-pr} that causes $\delta n_i \ne 0$ at the
ESP.\footnote{In this paper we will not discuss the mechanism of
inhomogeneous preheating.
See Ref.\cite{IH-pr} for the origin of the inhomogeneity $\delta n_i$ 
that appears in the number density.}
In the equilibrium phase of dissipative inflation, which is called ``warm
phase'' in this paper, the radiation temperature is determined
by $\phi$.
Therefore, $T_{eq}$ in Eq.(\ref{n-th}) is determined by the value of 
$\phi$ at that time.
Considering modest assumption that the evolution of $\phi$ is negligible
in the thermal phase, $T_{eq}$ is determined by $\phi_*$.
Therefore, $\delta \phi_*\ne 0$ may source both $\delta T_*
\ne 0$ and $\delta T_{eq}\ne 0$.  

In section \ref{multichi} we will consider an optional scenario in which
the dissipation 
coefficient has a jump in the warm phase.
We will show that the jump may appear naturally from multi-field 
interactions of the spot.
If the jump does not occur simultaneous in space, it
may cause creation of the curvature perturbation.

Before discussing our new scenario of dissipative inflation,
we first make a short review of the conventional warm inflation scenario,
mentioning the dynamics in the non-equilibrium phase where 
(in contrast to conventional warm inflation scenario) radiation
scales as $\rho_R \propto a^{-4}$. 

\subsection{Dissipative inflation and warm inflation (review)}
\label{diss-slo}
Warm inflation is the inflation scenario that uses dissipation for the
slow-roll\cite{warm-inflation-original}, but it has been used mainly for the
warm phase.
In the usual warm inflation scenario, radiation
density $\rho_R$ is sustained by continuous sourcing 
from the dissipation.
The sourcing mechanism works when thermalization is quick for the decay
products. 
As the result, $\dot{\rho}_R \ll 4H\rho_R$ is achieved
and $\rho_R$ in the warm phase is kept constant during inflation.
What discriminate our scenario of spotted inflation from 
usual warm inflation scenario are 
(1) origin of the initial radiation, (2) appearance of the thermal
phase\footnote{The thermal phase is first discussed in
Ref.\cite{Dissipative-curvatons}  for the dissipative curvatons. },
and (3) characteristic interactions of the spot.

Following the standard convention of warm inflation, the equation of motion
when the inflaton is dissipative is given by 
\begin{equation}
\ddot{\phi}+3H(1+r)\dot{\phi}+V(\phi,T)_{\phi}=0,
\end{equation}
where the subscript denotes the derivative with respect to the field
$\phi$.
We think one intuitively knows that the time-dependent mass of the
intermediate field $\chi_i$ causes excitation of $\chi_i$ particles,
which may lead to energy loss if $\chi_i$ particles decay into light
fermions. 
Of course the dissipation is not a trivial issue, but one immediately
 finds that the energy loss caused by the dissipation should appear as
 frictional force in the field equation. 
Understanding the dissipative dynamics from the quantum field theory
is a challenging issue.
(Note that the issue related to quick thermalization that sustains
radiation is basically independent from the 
dissipation mechanism. Namely, dissipation may occur without
thermalization of the decay products.) 
Much work has been done in this field. 
Analytic approximations are studied in Ref.\cite{Gamma-Ms0,
Gamma-Ms1, warm-analytic}, and numerical based methods are considered
in Ref.\cite{numerical-based}.
Quasiparticle approximation, which explains creation of radiation
in the zero-temperature Universe, is discussed in
Ref. \cite{Gamma-Ms0,Gamma-Ms1}. 
Dissipation in the background with non-zero temperature is discussed in
Ref.\cite{warm-analytic}. 
The latter is useful for spotted
inflation, while quasiparticle approximation is disregarded in this
paper because the contribution is less effective for the 
model.
The situation is in contrast to the model
considered in Ref.\cite{Dissipative-curvatons}.
Although the dissipation is well studied in terms of analytic and
numerical methods, quick thermalization is usually an assumption, because
thermalization, which is important for realizing the warm phase, is
determined by the interactions in the radiation sector, which is
arbitrary for the inflationary model.
Details for the thermalization are found in
Ref.\cite{thermalization-paper}. 
The strength of the dissipation is usually measured using $r$.
This rate is defined using the dissipation coefficient $\Upsilon$ and
the Hubble parameter $H$ as;
\begin{equation}
r\equiv \frac{\Upsilon}{3H}.
\end{equation}
The standard calculation of the dissipation coefficient $\Upsilon$
is mentioned later in this section.
The dissipation coefficient $\Upsilon$ is caused by
the microscopic physics associated with the interactions.
Then the energy conservation equation leads to;
\begin{equation}
\label{cons-rad}
\dot{\rho}_R+4H \rho_R = \Upsilon \dot{\phi}^2,
\end{equation}
where the right hand side ($\Upsilon\dot{\phi}^2$) represents
the source from the dissipation.
The radiation energy scales as $\rho_R\propto a^{-4}$ 
when the source is negligible, while $\rho_R$ behaves like a constant
($\dot{\rho}_R\simeq 0$)  
when the source is significant ($4H\rho_R \sim \Upsilon
 \dot{\phi}_{1I}^2$).
As we mentioned previously, we will denote the former by 
``thermal phase''.
The thermal phase is important when dissipative inflation starts with
non-equilibrium temperature\cite{Dissipative-curvatons}.

Scenario with $r>1$ is called strongly dissipating (SD) scenario.
In this paper we consider SD motion as we have in mind the situation
that dissipation controls the field motion during inflation.
Effective slow-roll parameters for the dissipative motion are basically
different from the conventional ones.  
They are defined as;
\begin{eqnarray}
\epsilon_w &\equiv& \frac{\epsilon}{(1+r)^2},\nonumber\\
\eta_w &\equiv& \frac{\eta}{(1+r)^2},
\end{eqnarray}
where the standard slow-roll parameters ($\epsilon$ and $\eta$)
are defined as;
\begin{eqnarray}
\epsilon&\equiv& \frac{M_p^2}{2}\left(\frac{V_\phi}{\rho_{Tot}}\right)^2,
\nonumber\\
\eta &\equiv& M_p^2\frac{V_{\phi\phi}}{\rho_{Tot}},
\end{eqnarray}
where $M_p\equiv (8\pi G)^{-1/2}\simeq 2.4\times 10^{18}$GeV is the
Planck mass and $\rho_{Tot}$ is the total energy density.
Before preheating, we assume the cold Universe.
Then, preheating leads to creation of radiation, 
which leads to significant dissipation for the inflaton motion.

The warm-phase condition is $4H\rho_R \simeq \Upsilon \dot{\phi}^2$.
When this condition is combined with slow-roll equations, it 
leads to more stringent conditions:
\begin{eqnarray}
\label{slo-ro-warm}
\epsilon &<& (1+r)\nonumber\\
\eta &<& (1+r).
\end{eqnarray}
These are the usual slow-roll conditions for the standard warm inflation
scenario\cite{gil-berera}. 
In this paper, the above conditions are used in the warm phase.

For the slow-roll condition in the thermal phase, where the warm-phase
condition is violated, we introduce an alternative $\dot{V}< 4H V$,
which suggests that the potential decreases slower than the radiation.

Before closing this short review, it
is useful to mention the origin of the dissipation coefficient
$\Upsilon$.
Following past results, it is useful to consider typical
superpotential used in Ref.\cite{Gamma-Ms1, Moss:2006gt}.
The superpotential for the interactions is given by
\begin{equation}
W=g_1\Phi X^2+g_2 XY^2,
\end{equation}
where $g_1$ and $g_2$ are coupling constants, and $\Phi, X, Y$ are
superfields whose scalar components are given by $\phi, \chi, y$, and
their fermionic partners are $\psi_\phi, \psi_\chi, \psi_y$, respectively.
In the dissipation at $T >H$, which is
important in our model, the dissipation coefficient usually depends on
the temperature $T$ and $\phi$.
The motion ($\dot{\phi}\ne 0$) causes excitation of heavy
intermediate field $\chi$ since the field $\chi$ has $\phi$-dependent mass.
Then dissipation occurs when the excitation of $\chi$ decays into light
fermion $\psi$.
The interactions are the same as the spot interactions,
except for the shift introduced for the inflaton field:
$\phi\rightarrow \phi-v_i$.

Considering the interactions of the spot, the mediating field is
$\chi_i$, which obtains mass $m_{\chi_i}\sim g_i |v_i-\phi|$.
The dissipation is caused by the excitation of the
mediating field $\chi_i$ that decays into light fermions.
Warm phase may not appear if thermalization delays.
The dissipation coefficient is given by $\Upsilon \propto T$ at high
temperature ($m_\chi \ll T$), while at low temperature ($m_\chi \gg T$)
it leads to $\Upsilon \propto T^3/|\phi-v_i|^2$.
In this section we consider the dissipation coefficient
that is given by
\begin{equation}
\Upsilon_i \equiv C_i \frac{T^3}{\Delta \phi ^2},
\end{equation}
where $\Delta \phi\equiv |\phi-v_i|$ denotes the distance from the ESP
and $C_i$ is dimensionless constant which is determined by the
couplings and the number of fields associated with the dissipation.
The above coefficient is used in both the thermal and the warm phases.
We implicitly assume that the temperature of the radiation always
satisfies $m_\chi > T$.

\subsection{Preheating at $\phi=v_i$ and creation of radiation at $\phi=\phi_*$}
Near the ESP at $\phi=v_i$, $\chi_i$ particles are effectively 
massless.
When the field $\phi$ passes through the ESP at $v_i$, the $\chi_i$
particles  are produced with the number density $n_i$, which is
maximally\cite{PR-original} 
\begin{equation}
n_i^{Max} \sim \frac{g_i^{3/2}}{(2\pi)^3}
\left(\dot{\phi}_0\right)^{3/2},
\end{equation}
where the initial velocity $\dot{\phi}_0$ depends on the scenario
before the $\phi$ field hits the ESP.
After creation of $\chi_i$ particles, they gain mass 
$m_i\sim g_i \Delta \phi$ as the field $\phi$ rolls away from the ESP.
Then, they decay into light fermions.
The decay occurs at $\phi_*$, where the decay rate of the $\chi_i$
particles becomes the same order as the Hubble parameter.
Since the decay rate is given by
\begin{equation}
\Gamma_i(\chi_i\rightarrow 2\psi)\simeq 
\frac{h_i^2}{8\pi} m_{\chi_i}\simeq 
\frac{h_i^2}{8\pi} g_i |\phi-v_i|,  
\end{equation}
the decay occurs at
\begin{equation}
\phi_*\simeq v_i-\frac{8\pi}{h_i^2 g_i}  H.
\end{equation}
We assume that the decay products quickly thermalize to raise radiation
just after $\phi=\phi_*$.
The radiation energy density at $\phi=\phi_*$ is thus given by
\begin{equation}
\label{Rstar}
\rho_R^*\simeq \Delta \phi n_i \simeq \frac{8\pi}{h_i^2 g_i}  H
\frac{g_i^{3/2}}{(2\pi)^3}
\left(\dot{\phi}_0\right)^{3/2}.
\end{equation}
Considering the slow-roll condition $\epsilon,\eta < (r+1)$ 
for the warm phase, the lowest temperature that is required to initiate
warm inflation  (i.e., the condition required for the dissipative
slow-roll in the warm phase) is given by  
\begin{equation}
\label{Tstar}
T> T_{min}\equiv \left[
\frac{Max[\epsilon,\eta] \times 3H(\Delta \phi)^2}{C_i}
\right]^{1/3}\simeq 
\left.
\left(\frac{Max[\epsilon,\eta] \times 64\pi^2}{C_i h_i^4 g_i^2}\right)^{1/3}H
\right|_{\phi=\phi_*}.
\end{equation}
Defining $\rho_R \equiv C_R T^4$ and using Eq.(\ref{Rstar}) and 
Eq.(\ref{Tstar}), we find the condition for the initial velocity at
$\phi=v_i$, which is given by
\begin{equation}
\dot{\phi}_0> \dot{\phi}_{min}\equiv 
\frac{C_R^{2/3} T_{min}^{8/3}h_i^{4/3}\pi^{4/3}}
{g_i^{1/3}H^{2/3}},
\end{equation}
which does not lead to significant constraint for the model.

\subsection{Dissipative inflation in the thermal phase}
Considering the equilibrium (warm-phase) condition 
$4H\rho_R \simeq \Upsilon \dot{\phi}^2$,
the temperature in the equilibrium phase of warm inflation is given by 
\begin{equation}
\label{equi-gen}
T_{eq}^7\simeq \frac{1}{4HC_R C_i}(\Delta \phi V_\phi)^2.
\end{equation}
Therefore, the Universe is in the thermal phase 
during $T_*>T>T_{eq}$.

Since  in the thermal phase the source term $\Upsilon\dot{\phi}^2$
 in Eq.(\ref{cons-rad}) is small compared with other terms,
the equation for the radiation energy density is given by
\begin{equation}
\dot{\rho}_R=-4H \rho_R,
\end{equation}
which leads to $\rho_R \propto a^{-4}$.
The number of e-foldings during this period is thus given by
\begin{equation}
N_{(th)}\simeq \ln \left(\frac{T_*}{T_{eq}}\right).
\end{equation}

Considering the inhomogeneous preheating scenario \cite{IH-pr} that
leads to $\delta n_i$ at the ESP, we find the cosmological perturbation
created at the beginning of the thermal phase;
\begin{equation}
\delta N^{ini}_{(th)} \simeq \frac{\delta T_*}{T_*}\simeq
\frac{1}{4}\frac{\delta n_i}{n_i},
\end{equation}
where $\delta n_i$ depends on the model considered for the inhomogeneous
preheating.
Since the slow-roll is significant and the time spent in the
thermal phase is very short, we may assume that at the end of
the thermal phase the inflaton field is still very close to
$\phi=\phi_*$.
Therefore, $T_{eq}$ in the above equation is estimated at
$\phi\simeq\phi_*$, and the inhomogeneity $\delta T_{eq}$ may be caused
by the modulation $\delta \phi_*$.\footnote{$\delta \phi_k$ at the
horizon exit is 
not relevant for the modulation $\delta \phi_*$.}  
In fact, in the equilibrium phase of warm inflation there is a direct
relation between the equilibrium temperature $T_{eq}$ and the field
value $\phi$.  
As the result, assuming $\phi\simeq \phi_*$ at the end of the thermal
phase, $\delta N$ caused by the end-boundary of the
thermal phase is given by
\begin{equation}
\delta N^{end}_{(th)} \simeq -\frac{\delta T_{eq}}{T_{eq}}\simeq
-\frac{\partial T_{eq}/\partial \phi|_{\phi=\phi_*}}
{T_{eq}}\delta \phi_*+O(\delta \phi_*^2),
\end{equation}
where $\delta \phi_*$ is not the amplitude of the pure(Gaussian)
perturbation of the field $\phi$, but is the modulation induced by other
light fields.
To understand the situation, we consider the simplest case
with monomial potential given by 
\begin{equation}
V(\phi)=\lambda_n \frac{\phi^n}{M_p^{n-4}},
\end{equation}
where the less effective terms are disregarded.
Using this potential, we find the equilibrium temperature
\begin{eqnarray}
T_{eq}^7&\simeq& \frac{1}{4HC_R C_i}(\Delta \phi V_\phi)^2\\
&\simeq & \frac{1}{4HC_R C_i}\left(\frac{\Delta \phi}{\phi}\right)^2
(n V)^2.
\end{eqnarray}
The temperature in the warm phase of the spotted inflation has a
significant suppression factor $(\Delta \phi/\phi)^{2/7}$, when it is
compared with the standard warm inflation scenario with the ESP
at the origin.

For the modulation $\delta h_i \ne 0$, the equation leads to 
\begin{eqnarray}
\delta N^{end}_{(th)} &\simeq&
-\frac{2}{7}\left[\frac{(n-1)}{\phi_*}
 -\frac{1}{\Delta \phi}\right]\delta \phi_*
\simeq \frac{2}{7}\frac{\delta \phi_*}{\Delta \phi}
\sim -\frac{4}{7}\frac{\delta h_i}{h_i},
\end{eqnarray}
where $\phi\simeq \phi_*$ is assumed.

\subsection{Dissipative inflation in the warm phase}
Obviously, the modulation associated with $\delta \phi_*$ may also cause 
$\delta N$ in the warm phase.
$\delta N$ caused by $\delta \phi_*$, which determines the perturbation
of the initial boundary of the warm phase, is given by
\begin{equation}
\delta N_{(warm)}^{ini}\simeq
\frac{\partial N_{(warm)}}{\partial \phi_*}\delta \phi_*+O(\delta \phi_*^2),
\end{equation}
where $N_{(warm)}$ depends crucially on the potential and the model
parameters.
However, introducing the inflaton velocity $\dot{\phi}_*$, which is
defined at $\phi=\phi_*$, the cosmological perturbation is expressed
by the simple form; 
\begin{equation}
\delta N_{(warm)}^{ini}\simeq -H\frac{\delta \phi_*}{\dot{\phi}_*}
\sim  -
\frac{16\pi H^2}{h_i^2g_i \dot{\phi}_*}\frac{\delta h_i}{h_i}.
\end{equation}
The total perturbation created by the modulation $\delta \phi_*\ne 0$
is given by the sum of $\delta N_{(warm)}^{ini}$ and $\delta N^{end}_{(th)}$.

The end of the warm phase is determined by the slow-roll condition
$\epsilon,\eta < (1+r)$.
For the monomial potential, it gives
\begin{equation}
\frac{v_i-\phi_e}{\phi_e} \simeq \left(\frac{M_p^4}{V}\right)^{1/8}
n^{3/4}C_i^{1/2}C_R^{-3/8},
\end{equation}
where strong dissipation and quick thermalization is assumed till the
end of warm inflation. 
Since the modulation of the coupling $h_i$ causes modulation of $C_i$
in the dissipation coefficient, the end-boundary of the warm inflation
can be modulated.\footnote{If one considers modulation in non-dissipative
inflation, one immediately understands that the interactions required
for the strong modulation tend to cause steep potential for the moduli.
In Ref.\cite{At-the-end-kov} a successful model is
constructed for the hybrid-type potential, however the modulation is
still a challenging issue in this field. In this sense, dissipative
inflation is an useful exception, which offers many kinds of modulation.}

If the warm phase lasts long ($N_e >60$) and the horizon exit occurs in
the warm phase, the curvature perturbation generated by the inflaton
$\phi$ is calculated precisely the same way as the standard warm
inflation scenario.
The only difference that may discriminate spotted inflation from
standard warm inflation is the shift of the ESP.

When the dissipation coefficient is independent of $T$, the amplitude of
the spectrum in the strong dissipative regime of the 
standard warm inflation scenario is given by\cite{spectra-warm}
\begin{equation}
{\cal P}_R^{1/2}\simeq \frac{H}{|\dot{\phi}|}
\left[
\left(\frac{\pi r}{4}\right)^{1/4}\sqrt{T H}
\right]
\simeq 
\frac{H^{5/4}}{3^{1/4}|\dot{\phi}|}
\left(\frac{\pi}{4}\right)^{1/4}\Upsilon^{1/4}T^{1/2}.
\end{equation}
On the other hand, when $\Upsilon$ depends on $T$, the temperature dependence of the dissipation coefficient can be
parameterized by 
\begin{equation}
c\equiv \frac{T \Upsilon_T}{\Upsilon},
\end{equation}
which shows that $\Upsilon \propto T^c$.
Using this parameter, the amplitude of the spectrum is given
by\cite{T-dep-spectra}
\begin{equation}
{\cal P}_R^{1/2}\simeq 
\frac{H^{3/2}T^{1/2}}{|\dot{\phi}|}f(r)^{1/2},
\end{equation}
where the function is $f(r)\sim A_c r^{3c+1/2}+B_c r^{3c-1/2}$
for $c>0$.\footnote{See Ref.\cite{T-dep-spectra} for the coefficients
$A_c$ and $B_c$. One can find more details of the result in
ref.\cite{T-dep-spectra}.}


Since the temperature in the warm phase is determined by $\phi$, 
the dissipation coefficients in the warm phase can be expressed using
$\phi$.
For the dissipation coefficient for the conventional warm inflation,
it leads to
\begin{eqnarray}
\Upsilon_0 &=&
 \left[\frac{V_\phi^2}{4HC_R}\right]^{3/7}C_i^{4/7}\phi^{-8/7},
\end{eqnarray}
while for the spotted inflation, it gives
\begin{eqnarray}
\Upsilon &=& 
 C_i \left[\frac{(\Delta \phi)^2 V_\phi^2}{4HC_RC_i}\right]^{3/7}
\left(\frac{1}{\Delta \phi}\right)^{2}\nonumber\\
&\sim & \Upsilon_0 \left(\frac{\phi}{\Delta \phi}\right)^{8/7}.
\end{eqnarray}
Therefore, even if the inflaton potential is identical to conventional
warm inflation, the ratio $r$ and the parameter $\beta$ are different in
spotted inflation.
The difference is due to the shifted interactions of the ESP.
The difference appears in the scalar perturbation spectra and the
spectral index, as well as in the equation for the number of
e-foldings. 
As the result, cosmological perturbations generated during inflation are
quite different in spotted inflation, even if the perturbations are
generated in the warm phase.

\subsection{Modulation of the dissipation coefficient during inflation}
In the above scenarios we considered modulation 
($\delta \phi_*\ne 0$)
and inhomogeneous preheating 
($\delta n_i\ne 0$)
to calculate the cosmological perturbation created during 
dissipative inflation.
We also considered conventional mechanism of generating curvature
perturbation during inflation.
In addition to these mechanisms, the modulation of the couplings (e.g.,
$\delta h_i$) may also cause 
inhomogeneity of the dissipation coefficient $\Upsilon_i$ during
inflation,
which leads to inhomogeneity of the inflaton velocity.
For example, consider the typical dissipation coefficient for the
low-temperature interaction\cite{gil-berera}
\begin{equation}
\Upsilon_i \sim 0.1\times g_i^2 h_i^4 N_{\chi_i}N_{\psi}^2\frac{T^3}{\phi^2},
\end{equation}
where $N_{\chi_i}$ and $N_\psi$ denote the effective number of the
species, or a more general form
\begin{equation}
\Upsilon_i \sim F(T,\phi).
\end{equation}
The modulation $\delta \Upsilon_i\ne 0$ during warm inflation causes
modulation of the inflaton velocity\cite{IH-mod1}.
From the slow-roll equation of motion in the strongly dissipative
scenario, it is given by
\begin{equation}
\delta \dot{\phi}\simeq -\delta \left(\frac{V_\phi}{\Upsilon_i}\right)
\simeq \dot{\phi}\left(\frac{\delta \Upsilon_i}{\Upsilon_i}\right).
\end{equation}
Considering the modulation scenario of the inflaton
velocity\cite{IH-mod1}, the integration of the perturbation
during inflation leads to
\begin{equation}
\label{modu-ra}
\delta N_{\delta \dot{\phi}} \simeq \int 
\frac{\delta \dot{\phi}}{\dot{\phi}}dt
\sim \frac{1}{2}
\left(\frac{\delta \Upsilon_i}{\Upsilon_i}\right),
\end{equation}
where the result is not proportional to the number of e-folding
because the perturbation associated with the inflaton velocity decays as
$\delta \dot{\phi}\propto e^{-2Ht}$. 
The decay of the perturbation after horizon exit is caused by the
compensation mechanism due to the degree of freedom associated with the scalar
metric perturbation of the time coordinate.\footnote{It may be possible to
construct a model in which the velocity perturbation does not decay
after horizon exit\cite{IH-mod1}, however the model is not relevant for
the present case.} 

\subsection{Natural gap in the dissipative coefficient : Multi $\chi$ model}
\label{multichi}
The inflaton velocity may have significant gap
during inflation, as we have discussed in Section \ref{gapin} 
for the simple inflationary model with spots.
Then the modulated boundary at the end of the spot
may cause cosmological perturbations. 

Considering warm inflation scenario with $r>1$, significant gap of the
inflaton velocity may appear
when the effective number ($N_{\chi_i}$ or $N_{\psi}$) changes during
inflation. 
To construct a model with the gap in $N_{\chi_i}$, let us remember that 
 the effective mass of the intermediate particle $\chi_i$
is given by $m_i\simeq g_i\Delta \phi$, which must satisfy the condition
$m_i>m_\psi$ so that the decay $\chi_i \rightarrow 2\psi$ is possible 
for the dissipation.
In this case, the effective number of $\chi_i^{(j)}$ field is determined
by the condition $m_i^{(j)}>m_\psi$.
Therefore, if (1) there are two species $\chi^A_i$ and $\chi^B_i$ 
for the intermediate field that couples to $\phi$ with the interaction
$g_i^A\ne g_i^B$, and (2) the light fermion has bare mass $m_\psi\ne 0$,
there are two distinctive regions of dissipative motion 
in which the effective
number $N_{\chi_i}$ is given by either $N_{\chi_i}=1$ or $N_{\chi_i}=2$.
In fact, the dissipation caused by the intermediate field $\chi_i^A$ is
effective when 
\begin{equation}
g_i^A |\Delta \phi| \ge 2m_\psi,
\end{equation}
which is not identical to the region for the field $\chi_i^B$:
\begin{equation}
g_i^B |\Delta \phi| \ge 2m_\psi.
\end{equation}
The change of the number $N_{\chi_i}=1 \leftrightarrow  N_{\chi_i}=2$
 induces significant gap in the dissipation coefficient 
$\Delta \Upsilon_i /\Upsilon_i =1$, which leads to the required 
gap in the field equation.
Then, the modulation of the boundary between the two distinctive regions
causes cosmological perturbation due to the mechanism considered in
 Section \ref{gapin}. 

\section{Spotted curvatons}
\label{curv}
In the above scenarios we mainly considered the effect of the spot
during primary inflation.
In this section we examine the scenario in which the spot interaction
appears for the curvatons.
It is shown that the dissipative curvaton scenario with the spot
interaction is practically the scenario of secondary dissipative
inflation. 
Cosmological scenario of dissipative curvatons has already been
discussed in Ref.\cite{Dissipative-curvatons}.
The result obtained in Ref.\cite{Dissipative-curvatons} shows that
the dissipation coefficient derived using quasiparticle approximation
(zero-temperature approximation) dominates the field
equation of the dissipative curvaton when the ESP is placed at the origin.
The purpose of this section is to show why and how the
scenario of the dissipative curvaton becomes different in the spotted
model, in which the ESP is placed away from the origin.

We simply assume that the curvaton $\varphi$ is 
slow-rolling during primary inflation.
This assumption is the same as the conventional curvaton scenario.
Unlike the standard curvaton scenario, the potential of the curvaton is not
necessarily quadratic because the matter-like evolution during
oscillation period is not essential for the model.

We have in mind the situation that the curvaton energy density is
negligible during inflation but its ratio grows during dissipative
slow-rolling.
The situation is in contrast to the usual curvaton scenario, in which
the ratio grows during oscillation.
It would be possible to construct many different scenarios in this
direction, however the most useful scenario would be that curvaton
oscillation starts when $H=H_{osc}$ and then the curvaton passes through
the ESP at $\varphi = v$. 
Our modest assumption is that the radiation caused by the primary
reheating has significant interactions with the spot on the curvaton
trajectory. 
Then, using the dissipation in the non-zero temperature background, 
the dissipation coefficient of the curvaton motion is enhanced near the
ESP, which stops the curvaton motion at the first oscillation.
At first the dissipative motion is in the thermal phase, 
and then the warm phase may appear if the dissipation leads to quick
thermalization of the decay products.
Anyway, what is important for the curvaton scenario is whether the
curvaton can start dominating the energy of the Universe before it
decays into radiation.
If the dissipation coefficient associated with the spot is given by
\begin{equation}
\Upsilon_\varphi =C_\varphi\frac{T^3}{|\varphi-v|^2},
\end{equation}
the slow-roll parameters in the thermal phase lead to
\begin{eqnarray}
\eta < (r+1)^2 &\rightarrow& T^3>
\frac{\eta^{1/2} H |\varphi-v|^2}{C_\varphi}\\
\dot{V}<4HV \sim \epsilon < (1+r) &\rightarrow& 
T^3 > \frac{\epsilon H |\varphi-v|^2}{C_\varphi}.
\end{eqnarray}
These equations show that domination by the curvaton potential is indeed
possible 
even if the warm phase does not appear for the dissipative curvaton.
The number of e-foldings elapsed during dissipative inflation, which is
 associated with the curvaton field $\varphi$, is controlled by the
 initial value of $\varphi$ and the spot interactions.
Since the analyses related to spotted curvatons are highly
model-dependent, we will not discuss this issue further in this paper.

\section{Conclusions and discussions}
In this paper we described new scenarios of inflation and mechanisms of
 generating curvature perturbations associated with the spot
 interactions. 

Our first example uses a velocity gap that does not occur simultaneous
in space. 
The best known example of this kind is the scenario of generating 
cosmological perturbations at the end of hybrid
 inflation\cite{At-the-end-lyth,At-the-end-kov}, in which 
inflaton velocity changes suddenly at $\phi=\phi_e$ due to the
 instability caused by the waterfall field.
In this case, additional inflaton field\cite{At-the-end-lyth}
or modulation\cite{At-the-end-kov} causes $\delta \phi_e \ne 0$, which
leads to inhomogeneous transition associated with the gap in the velocity.
We generalized this idea by introducing a gap 
caused by the spot interactions.

Considering the spot interactions, similar interactions
appear in the trapped inflation scenario.
In the trapped scenario, oscillation around the ESP(trapping) occurs
because the preheat field $\chi_i$ does not decay. 
In this sense, our scenario complements the trapped scenario,
considering the decay of the preheat field. 

For the second example we considered dissipative inflation caused by the spot
interactions. 
We specifically considered the situation in which the Universe is cold
before the preheating, and then significant radiation is created by the
spot interactions. 
The model can be used to construct a two-step inflation scenario, in which
the Universe is initially cold but warm inflation starts in the midst of
the inflation epoch. 

For the third example we considered spotted curvatons, in which the
two-step scenario is extended using two independent fields: inflaton
$\phi$ and the curvaton $\varphi$.

In this paper we focused our attention on showing existence proofs of these
peculiar scenarios.
In conventional models of inflation and curvatons, significant
interactions are usually troublesome.
Therefore, inflaton or curvatons are
usually isolated from the standard model (SM) sector.
In fact, constructing cosmological models with significant
 interactions is a formidable task\cite{IH-mod1,
matsuda-warm, matsuda-warm-apps, formida-ad, Topolo-curv, Hybrid-curvatons}.
Strong interactions of the inflaton field may cause high reheat
temperature, which may lead to significant production of unwanted stable
or quasi-stable relics.
For the curvatons, strong interactions may ruin longevity of the curvaton 
oscillation\cite{long-MSSM}, which is essential for the original
curvaton model\cite{curvaton-paper, matsuda_curvaton}.
In this sense, our examples of spotted inflation and spotted curvatons
may seem quite peculiar.
However, we believe that these models are important for
developing cosmological models with significant interactions,
which may be useful to find the scenario without isolated sector that
may sometimes be introduced only to explain the specific cosmological scenario.
Moreover, modulated scenarios for generating cosmological perturbations, 
which are considered in the paper, are very useful when dissipation is
significant.\footnote{As we already mentioned in the text, 
interactions required for the strong modulation sometimes raises the
moduli potential and may conflict with the massless condition when the
modulation is considered for the inflaton potential\cite{At-the-end-kov,
IH-mod1}.}
We believe there are useful possibilities of the scenario,
especially when it is applied to particle models of the grand
unification or the string theory.
However, since our task in this paper is not to show specific
applications of the scenario, they will be discussed elsewhere
in separate publications.

\section{Acknowledgment}
We wish to thank K.Shima for encouragement, and our colleagues at
Tokyo University for their kind hospitality.

\end{document}